# Smarter AI Through Prompt Engineering: Insights and Case Studies from Data Science Application

Snehasish Paul[1*], Rohit Kumar[2] and Laxman Das[3]

[1,2] Department of Library and Information Science, University of Delhi, Delhi, Pin- 110007, India

[3] Symbiosis Institute of Business Management, Noida, Symbiosis International (Deemed University), Pune, Pin-201301 India

**\*Corresponding Author-** snehasishpaulas98@gmail.com

**ORCID ID-** https://orcid.org/0009-0003-2730-5314

**Abstract**

The field of prompt engineering is becoming an essential phenomenon in artificial intelligence. It is altering how data scientists interact with large language models (LLMs) for analytics applications. This research paper shares empirical results from different studies on prompt engineering with regards to its methodology, effectiveness, and applications. Through case studies in healthcare, materials science, financial services, and business intelligence, we demonstrate how the use of structured prompting techniques can improve performance on a range of tasks by between 6% and more than 30%. The effectiveness of prompts relies on their complexity, according to our findings. Further, model architecture and optimisation strategy also depend on these factors as well. We also found promise in advanced frameworks such as chain-of-thought reasoning and automatic optimisers. The proof indicates that prompt engineering allows access to strong AI localisation. Nonetheless, there is plenty of information regarding standardisation, interpretability and the ethical use of AI.

## Introduction

Large language models are transforming how computers understand and communicate in human language. Previously, AI performance was improved by modifying architecture, extensive fine-tuning, and training them in a specific domain. Because of all these complications, it took a lot of computing and know-how to do this in practice. The broad language models obtain much knowledge and reasoning mechanisms during pre-training. The engineering input design reveals the knowledge and reasoning mechanisms that are embedded in the patterns of these convolutional neural networks. Recent techniques never consider a prompt as a simple question; rather, it is an interface through which the model can concentrate on a particular set of knowledge while constraining the output space for the expected tasks. Researchers have proposed various methods, including zero and few-shot learning, chain of thought, retrieval enhanced learning and automated optimization framework, due to this insight. In a region, data use offers ample opportunities to innovate through prompt engineering (Ge et al., 2025). This is because it offers a multitude of analytic tasks. In addition, it is often domain-specific with complex requirements. Moreover, traditional limitations have stopped non-technical stakeholders from leveraging AI usage. Throughout the years, the process of transforming raw data into sense data often required high levels of skill with programming and subject matter knowledge. With prompt engineering, one is able to give natural language specifications of analytical goals, thereby making a larger number of people capable of playing a meaningful role in the data science workflow. Currently, a project is being undertaken that examines the various prompt engineering techniques and which ones work better in which data science contexts. Similarly, it will look at which type of prompts are most useful for which type of analysis. Lastly, it will study which steepest descent or other forms of optimisation framework gives the best trade-off between performance and complexity.

**Methodological Foundations of Prompt Engineering**

Prompt engineering today includes many techniques and these differ according to sophistication, computing requirements and contexts. The reasoning prompt is a good addition to the approach used for prompt engineering. Chain of Thought Prompting (CoTP) is when we add reasoning structures on top of existing models to engage in multi-step problem-solving processes. CoTP means logical pathways branch out, allowing for other solution strategies to be pursued simultaneously along with the initial one. Generally, a framework that enables structured reasoning is useful for completing a task that requires higher level reasoning or maths. According to Khatuya et al. (2025), their FINDER framework shows improvements of 5.98 and 4.05 percent in the accuracy of the execution of the FinQA and ConvFinQA benchmark datasets, respectively, by combining the FINDER retrieval and programmatic prompting. It shows that the integration of reasoning structures directly into the design of the model can improve the performance of the model for tasks that require logical sequence.

Instruction prompts detail the specific task and format of the output. These prompts use direct language to express desired behaviours and limit output specifications. Hu et al. (2024) provided evidence of the usefulness of structured instruction prompts in clinical named entity recognition tasks by developing a four-component framework made up of baseline task descriptions, instructions based on the guidelines, instructions for error analysis, and annotated few-shot examples. The findings showed that under the systematic application of all prompt components, GPT-4 Relaxed F1 Scores started from 0.804 and rose to 0.861 on the MTSamples datasets while on the VAERS datasets it rose from 0.593 to 0.736. The considerable improvements show that a good instruction design should consider a multiple aspect of specification at the exact moment.

Contextual prompts highlight the importance of providing models with information on the task at hand as well as the wider context, domain knowledge, and examples. In some disciplines where the right answer relies on the technical vocabulary, culture or historical background, this technique is useful. According to Polak & Morgan (2024), conversation prompts were used in the ChatExtract system to extract materials data. The values of accuracy and recall metric of the material properties in advanced conversational models were nearly 90 percent. Using rich contextual information that can be verified iteratively can improve extraction in engineering domain, as shown by this method.

Using advanced automated optimization systems enables the systematic and google search of various prompt design spaces for the outcome of RNNs. Lieander et al. (2025) presented PO2G, a gradient based optimization mechanism, which stands for 'Prompt Optimization with Two Gradients'. After a total of three iterations, method achieves approximately 89% accuracy. This is better than ProTeGi baseline which needed six iterations for same performance. We can optimize algorithms so that work on the prompts can be done faster than humans are able to do. According to Agarwal et al. (2024), PromptWizard is an agent-driven framework that provides iterative mutation of instructions and examples along with critic-based evaluation. It was found that over 35 evaluation tasks, PromptWizard consistently demonstrates improvement over existing prompt strategies.

**Comparative Effectiveness Across AI Models and Tasks**

Prompt engineering techniques can be effective for some model architectures and tasks but less so for others. It's important to understand performance variations for practitioners aiming to choose a model and a prompting strategy for a specific use. Recent studies help us gain insight into these relationships through empirical evaluation across an increasing number of models and benchmarks.

Table 1: Key Model Performance Indicators from Real-World Prompt Engineering Research

| Model | Task Domain | Performance Metric | Source |
| --- | --- | --- | --- |
| GPT-4 | Clinical NER | Relaxed F1: 0.861 (MTSamples), 0.736 (VAERS) | (Hu et al., 2024) |
| GPT-3.5-turbo | Job Classification | +6% Precision@95% Recall vs. supervised | (Clavié et al., 2023) |
| GPT-3.5-turbo / Claude 2 | Phishing Detection | F1: 92.74% (prompting approach) | (Trad & Chehab, 2024) |
| GPT-4 | Data Preprocessing | 100% accuracy/F1 on 4 of 12 datasets | (Zhang et al., 2023) |
| GPT-4 | Schema Matching | Recall: 100% (DeepMDatasets), 91.8% (Fabricated) | (Feng et al., 2024) |
| Claude | Structured Data Tasks | Highest accuracy across task set | (Schmidt et al., 2025) |
| ChatGPT-4o | Structured Generation | Most token- and time-efficient | (Schmidt et al., 2025) |

Table 1 shows selected performance highlights from empirical studies looking at how effective prompts are across models and tasks. According to the data, several trends emerge about model capabilities and prompt responsiveness.

Hu et al. (2024)'s clinical named entity recognition results show that neural networks such as GPT-4 respond more usefully to structured prompt components than their predecessors. The fact that the model gets substantial (F1) improvements when applying system prompt framework suggests that sophisticated models possess latent capabilities that can be activated via prompting mechanism. Yet a large performance gap between baseline and optimised prompts suggests that careful prompt design is essential for these capabilities rather than just an outcome of complex models.

As shown by Clavié et al. (2023),Carefully crafted zero-shot prompts can enable GPT-3.5-turbo to outperform a supervised learning approach on job classification tasks. In fact, it improved the Precision at 95% Recall by 6% compared to the best supervised baseline. These findings go against the common belief that task-specific training data is needed for a model to be successful. We find prompt engineering can act as a viable alternative to supervised learning pipelines in some cases. This has important practical implications as organisations may be able to deploy effective classification systems without the need to invest in labelled training data collection and model fine-tuning.

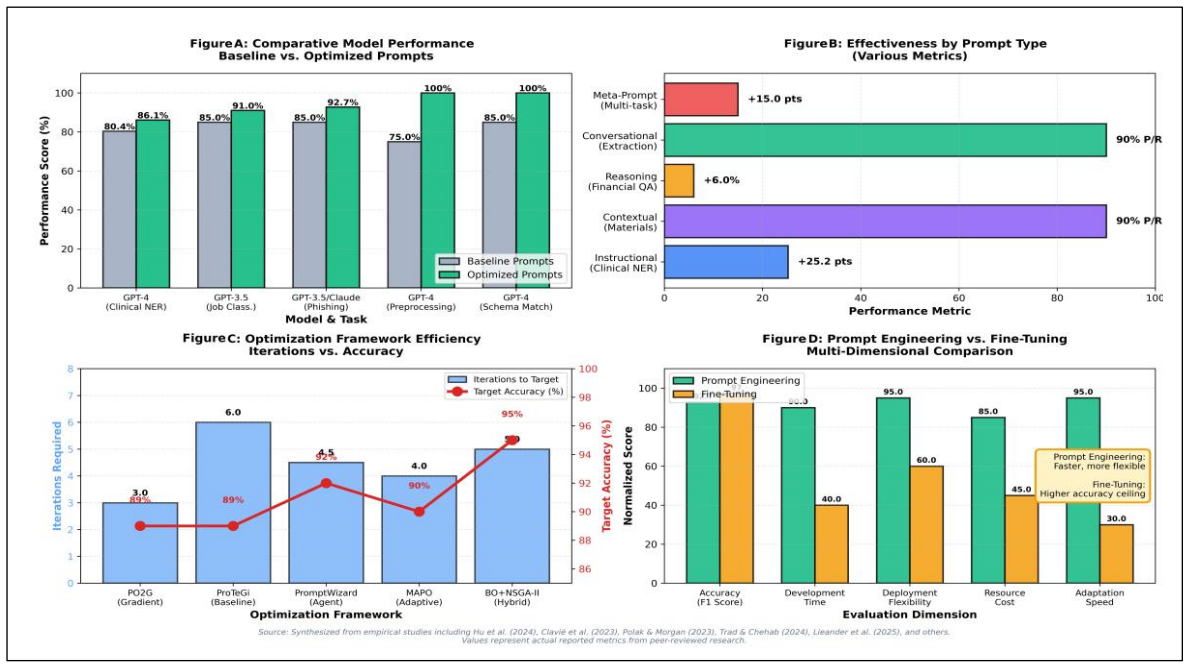

*Figure 1: Comprehensive Effectiveness Analysis of Prompt Engineering in Data Science Application*

Figure 1 Panel A presents the performance of baseline versus optimized prompts on five different model-task combinations, showing improvements between 5.6 to 25 percentage points. The contextual and conversational approaches are achieving ~90% precision/recall. Panel C shows the efficiency of the optimization framework. PO2G only needs 3 iterations, while baseline approaches need 6. Panel D compares prompt engineering and fine-tuning in multiple dimensions. In terms of accuracy, fine-tuning is superior (97.29% F1 vs. 92.74% F1). In terms of deployment flexibility and speed, prompt engineering has the edge.

(Schmidt et al., 2025) undertakes a comparative analysis of structured data generation by ChatGPT-4o, Claude, and Gemini. The study shows notable shifts in approaches regarding accuracy, efficiency, and resource use. The researchers found that Claude produced the most accurate output among the AI. Gemini was balanced across metrics while ChatGPT-4o was the best in tokens and time. According to the findings, choosing the right model is determined by real-world constraints and requirements. This is because different models have mismatched performance characteristics on different dimensions.

(Trad & Chehab, 2024) consider the pros and cons of prompt engineering versus fine-tuning in phishing detection are useful. The F1 scores of Claude 2 and GPT-3.5-turbo using prompt engineering test were found to be 92.74% and AUC values were 97.84%, while the fine-tuned models scored 97.29% and 99.56% on the same test set. The gap in performance suggests that even though deploying and using prompt engineering can happen quickly, fine-tuning might still be needed. This is applicable for applications that need utmost precision or for high-stakes use cases. In those cases, it is worth the additional effort to create a better product.

**Types of Prompt Engineering for Effective AI Responses**

The prompt engineering techniques can be classified into different categories based on their intention and working principle. When practitioners understand these categories, they will be able to pick appropriate approaches for particular analytical tasks. They will also be able to create combinations of different techniques to work together in an integrated workflow

The fundamental category Instructional Prompting focuses on giving clear and explicit instructions a computer program can use to do a specific task or procedure. Most these prompts use an imperative language format with structured layouts and output specifications to enhance clarity and reduce ambiguity. How well instructional prompts work depends largely on how precise the wording is, how instructions are logically organised, and how far outputs are constrained. (Hu et al., 2024) revealed that adding systematic guidance and error feedback to the entity recognition task substantially improves GPT-3.5's clinical performance on the task. When the suggestion was made less restrictive, there was an uptick of scores on MTSamples from 0.634 to 0.794.

Contextual prompting methods provide a rich background for AI to understand the context of a specific task or question. This approach is especially useful for domain-specific applications where specialized knowledge, technical terms, or cultural understanding matters for responses. According to (Polak & Morgan, 2024) this tool has demonstrated what is possible with contextual prompting for ChatGPT. By heavily utilizing domain-specific context, as well as conversational follow-up, they achieved 90% precision and recall scores for materials property extraction. The examples, historical context, and the use of technical terms in the replies improved their quality and relevance. We use the term technical because they used it in their uses.

Reasoning prompts help models improve their complex reasoning abilities by directly embedding reasoning knowledge into the prompts structure. Chain of thought prompts require explicit identification of intermediate steps of reasoning, whereas tree-of-thought approaches allow for the simultaneous exploration of multiple solution trees. These techniques are particularly effective for mathematical computations, logical reasoning and other multi-step analysis tasks. (Khatuya et al., 2025) show that continous instruction of cognitive behaviour together with flexible retrieval mechanisms can enhance the execution accuracy of financial question-answering benchmarks by approximately 6% on FinQA and 4% on ConvFinQA as compared to the baseline approaches.

To ensure there is continuity in conversation, we utilize Conversational Prompting strategies. Systems use dynamic management of context, personality consistency, and change of communication styles of language model, and other techniques. The success of conversation elicitation depends on the sophistication of the management and the retention of the Context during extended interactions. According to (Polak & Morgan, 2024), ChatExtract used systematic follow-up questions in its interview prompts in order to verify and improve the accuracy of raw material information.

When NLP systems create new prompts or change the original prompts to accomplish a task, this is called meta prompting. By using recursive techniques, we can increase the complexity of the optimization process. Moreover, one can obtain results that are usually more robust than those typically obtained from manual processes. MAPO fine-tunes the prompts for each LLM architecture. MAPO approach from (Chen et al., 2024). Specific model customisations have led to a steady increase in accuracy for several tasks.

Table 2: Characteristics and Effectiveness Metrics of Prompt Engineering Categories

| Category | Primary Applications | Key Characteristics | Reported Effectiveness Range |
| --- | --- | --- | --- |
| Instructional | Task specification, format control | Explicit directions, structured output | F1 improvements: 0.16-0.20 (clinical NER) |
| Contextual | Domain-specific tasks, specialized knowledge | Rich background, examples, terminology | Precision/Recall: ~90% (materials extraction) |
| Reasoning | Complex problem-solving, mathematical tasks | Sequential logic, intermediate steps | Accuracy gains: 4-6% (financial QA) |
| Conversational | Extended interactions, iterative refinement | Dynamic context, adaptive responses | Precision/Recall: ~90% (iterative extraction) |
| Meta-Prompting | Automated optimization, model-adaptive tuning | Self-refinement, model-specific adaptation | Consistent task improvements (MAPO study) |
| Optimization | Systematic search, efficiency improvement | Algorithmic refinement, multi-objective | 3-6 iteration reduction (PO2G vs. baselines) |

According to existing literature, Table 2 summarises the features, key usages, and reported effectiveness ranges of these major prompt engineering categories.

(Kusano et al., 2025) add an important nuance to our understanding of prompt effectiveness. Different prompting strategies work best depending on the model and its cost, best. They have performed extensive experimentation over 23 types of prompts, 12 LLMs, and various recommendation datasets. They find out that high-performance models usually work better with simpler prompts, and more cost-effective models benefit from complex reasoning prompts and contextual enhancements. The finding shows that prompt engineering must be tailored to the model and the job. Prompt engineering should take account of the model's capabilities and limitations.

**Optimisation Frameworks and Implementation Strategies**

The systematic optimization frameworks can support practitioners in efficiently exploring complex, flexible design space which can involve multiple performance criteria. Different frameworks to develop and enhance prompt engineering techniques are gradient-based approaches that estimate the surfaces of prompt effectiveness and evolution algorithms that sample discrete prompt spaces through iterative mutation and selection.

The gradient-based optimisation methods consider prompt refinement as a continuous optimisation problem that uses performance gradient approximations. According to research conducted by (Lieander et al., 2025) the PO2G framework developed by Lieander and his co-authors is pretty efficient. After three rounds of optimisation, their framework reached about 89% accuracy. Note that the ProTeGi baseline which is used for comparison reaches a similar percentage but only after six rounds of optimisation. As a result, they reduce the cost of computation and the time to deploy the algorithm. Thus, they are convincing options for resource-constrained applications.

The effectiveness of a prompt is fine-tuned with the help of a critic and a loop. PromptWizard was introduced by (Agarwal et al., 2024), an agent-based loop which evolves instruction and example with a critic-based evaluation in a loop. They evaluated across 35 different tasks and found to be systematically better than existing prompt strategies, suggesting that agent-driven approaches can find effective prompting patterns that are not easily discovered via manual design or through a simpler optimisation algorithm.

An important refinement is model-adaptive optimisation. This is the recognition that not all LLM architectures respond similarly to various prompting strategies. In (Chen et al., 2024), the authors developed MAPO, which designs prompts for specific models rather than universal prompt formulation. By adapting prompts to the special aspects of each model, task accuracy was improved consistently, indicating that optimising prompts to a model may get better results than optimising to the task.

Evolutionary and hybrid approaches utilize search techniques that employ a population-based method. In their 2025 research, (Hazman et al., 2025) utilized grammar-guided genetic programming as well as local search for discrete prompt optimization. They achieved performance benefits over the relevant baselines on several small LLMs and domain tasks. Efforts to design evolutionary approaches seem to suggest that the prompt design space contains a rich and complex structure that is most effective when searched using populations rather than through local search guided by gradients.

A multi-objective optimisation framework regards prompt effectiveness as evaluable along multiple dimensions at the same time. According to (Narayanaswamy & Muniswamy, 2025), NSGA-II collaborates with Bayesian optimisation to balance three objectives: accuracy, efficiency, and interpretability. Based on their framework, GLUE benchmark experiments obtained the following scores: 95 percent accuracy, 85 percent efficiency, and 79 percent interpretability. This indicates that systematic multi-objective optimisation can discover most configurations that perform well on concurrent evaluation criteria.

Cost-aware optimization recognizes that utilizing LLM through an API comes at a price, but its advantages must offset it. (Feng et al., 2024) regard schema matching as an NP-Hard optimisation problem in the presence of constraints imposed by GPT-4. They come up with ways to earn as much as possible from the matches at a cost. The prompt-matcher framework achieves a 100% and 91.8% recall on DeepMDatasets and Fabricated-Datasets, respectively, and explicitly regards the computational cost.

**Real-World Case Studies and Implementation Outcomes**

Different things show that it works well with many uses. Specifically, it is helpful within data science workflows. Also, it works on many real-world use cases. These cases illustrate the effectiveness of different prompting strategies and the expected performance advancements that may be achieved from these prompts in the field.

Polak & Morgan (2024) developed ChatExtract, a chat-based system for extracting information about material properties from literature. The method used systematic follow-up inquiries to check and refine extractions. Such processes of verification and mutation resulted in precision and recall results of 90% plus using top prompt models like ChatGPT-4. The system was able to successfully extract cooling rates and yield strengths from research papers. The output is done in a structured way in the database. It is done in a way that would have taken a lot of manual effort through traditional means of extraction. The case shows how a conversational prompt to refine followed by iterative verification enables highly accurate extractions in a technical task while saving much effort.

According to Clavié et al. (2023), job postings can be categorized as graduate jobs or entry-level jobs using prompt engineering, which is a form of business intelligence. They created a special zero-shot prompt for gpt-3.5-turbo which produced a precision at 95% recall improvement of 6% over the best supervised learning baseline. This is significant as it shows that prompt engineering can match performance of a supervised approach without having to collect training labels or fine-tune models. The study is likely to have some practical implications, such as a faster implementation timetable, lower costs for data collection and a quicker adaptation of the classification systems to changing labour market conditions.

Hu et al. (2024) suggested a novel framework that uses prompt engineering for clinical named entity recognition as mentioned in paper. The framework gives motivation for this framework. A careful evaluation of various task-specific contributions. When you use a complete prompt framework, your model performs much better on MTSamples and VAERS datasets. GPT-4 relaxed F1 score increased from 0.804 to 0.861 (MTSamples) and from 0.593 to 0.736 (VAERS). In clinical settings, accurate recognition of important entities can affect subsequent clinical decision-making and patient safety tools.

Trad & Chehab (2024) compare prompt engineering versus fine-tuning for phishing detection with GPT-3.5-turbo and Claude 2. Their prompt engineering methods achieve an F1 score of 92.74% on test set with 1.000 samples. This is quite good for a zero-shot method. However, these fine-tuned specialists models had F1 score of 97.29% with AUC value of 99.56% on the same test set, showing a big gap. This case study shows that although prompt engineering can allow AI models to be deployed more rapidly, and allows applications greater flexibility, applications that seek maximum accuracy, or that are security-sensitive, may still want to pursue model fine-tuning for further performance gain.

Feng et al. (2024) leveraged the abilities of GPT-4 with two specially designed prompts for semantic matching and abbreviation matching to tackle the schema matching problem. Using budget-aware selection algorithms to manage API costs, Their Prompt-Matcher framework achieves recall rates of 100% on DeepMDatasets and 91.8% on Fabricated-Datasets. The system formulated correspondence selection as an NP-Hard optimization problem under cost constraints and developed approximation algorithms to maximize matching quality whilst respecting budget limitations. This case shows how prompt engineering can be made to work with formal optimisation schemes to deal with real-world deployment constraints.

Zhang et al. (2023) evaluated the use of prompt engineering with LLMs for a set of basic data preprocessing tasks, which include error detection, imputation and entity matching. A comparative study conducted by Goldwasser et al. (2023) in their paper "Evaluating GPT-3.5, GPT-4, and Vicuna-13B on the WITH Dataset"" on 12 different datasets across GPT-3.5, GPT-4 and Vicuna-13B showed that GPT-4 got an accuracy or F1 score of 100% on 4 datasets with decent performance on some of the structured preprocessing tasks. Performance still differs greatly across different preprocessing operations, datasets and other aspects. It shows that prompt engineering requires careful design specific to task and validation for effective data preprocessing.

He et al. (2025) looked at human performance in prompt engineering for data labeling tasks in the absence of gold labels and found practical limitations. In their study, while it was found that unguided human prompt refinement could be challenging, some of the participants did improve with several iterations. However, it was noted that only 9 of the 20 displayed better labeling accuracy after four more iterations. This shows that a systematic optimization framework and automated prompts are important, to the extent that human intuition is insufficient to identify good prompts when clear feedback on performance is not available.

**Discussion and Practical Implications**

Experiments in different settings reveal that prompt engineering works and is feasible. Prompt engineering often depends on the task. This means that prompt engineering often gives small but still reliable gains (6% in some classification tasks) while in others the gain is sizeable (for example, 30 percentage points in some clinical NER applications). In other words, practitioners must test prompt engineering empirically for their use case and not assume it can be useful for all. The situation becomes less clear when, in the second case, one considers the relationship between sophistication of a model and the ideal prompting strategy. According to Kusano et al. (2025), high-quality models produce better results from simple prompts, while cheap models get better results from complex reasoning and context improvement. This means we ought to use prompt engineering techniques for the specific capabilities of the model and not across architectures. The selection of either prompt engineering or fine-tuning to achieve a desired performance is strongly influenced by the application's functionality, time to deployment and performance threshold requirements. According to Trad & Chehab (2024), prompt engineering leads to the accelerated deployment and solid performance of the model (F1 = 92.74%). However, fine-tuning might lead to meaningful additional gains (F1 = 97.29%) that may be necessary for critical applications. Companies must choose between prompt engineering, which is fast and flexible, and fine-tuning, which yields better performance. Systematic optimization frameworks can be beneficial to better manual prompt crafting is the fourth element. For instance, gradient-based and agent-based approaches lead to better performance and require less iterations. PO2G achieved the targeted performance in three iterations, while the baseline approaches needed six. It incurred lower computational costs and time-to-deploy Lieander et al. (2025). Investment in continuous optimization infrastructure could bring organizations deploying prompt-engineered systems at scale huge returns thanks to these efficiency gains. In practical areas where effective systems need to be assessed along more than one dimension simultaneously, multi-objective optimization frameworks explicitly demonstrating a trade-off or balance between accuracy, efficiency and/or interpretability are important. A hybrid approach of Bayesian and Genetic algorithms can be used to obtain prompt configurations that give an accuracy of 95%, efficiency of 85%, and interpretability of 79% by multi-objective optimisation systematically to give quite competitive results in complex multi-objective trade-off space.

**Limitations and Future Research Directions**

At the moment, prompt engineering research and practice deal with various limitations. Researchers make use of different metrics, benchmarks or experimental protocols. This limits systematic comparisons due to a lack of standardized assessment protocols. To achieve comparative authority and build on cumulated knowledge, more studies are needed for the development of evaluation frameworks. Changing the text from time to time in order to get an improved outcome presents a major difficulty that one faces. It is difficult to keep and change prompts because of that. It occurs more often in production environment when there may be more of a need of changing the prompt. Minor perturbations require strong prompting strategies that research must help us devise. Automated optimization frameworks are also hard to interpret. A complex optimization algorithm might find a prompt that yields good results, but we do not know why. If we could figure out why certain prompts perform well, we could learn to make prompts systematically. It would also help us when things don't work as expected. Cultural sensitivity and bias mitigation should be on our radar. A question useful in one culture or language might produce bad or biased answers in another. Exploring prompting strategies that are culturally adaptive and staying effective across diverse populations are important and ethical research areas. Merging prompt engineering and other techniques to enhance AI like RAG, use of tools, and multi-agent systems together is an exciting research direction. It may be that recognizing how prompt engineering relates to these synergistic skills may lead to results better than anyone technique achieves alone.

**Conclusion**

Prompt engineering enhances the functionality of AI in a wide range of data science applications. Also, it improves measurable effectiveness and reduces technical obstacles for AI adoption. Research indicates that prompting can improve performance anywhere from 6% to over 30%. The gain is improved by more complex models and optimizers. Using systematic optimization frameworks is more effective than spending time on human-curated prompts. One strategy achieves target performance in fewer steps through a gradient-based approach while agent-based approach may discover better-performing prompts. More than performance metrics, practical use is about making AI more powerful, speeding up time to market, and increasing flexibility to adapt to change. Nonetheless, one will not be able to obtain these benefits without careful attention to task optimization, model appropriate strategy selection, and multiparameter evaluation. As the area develops, standardized evaluation frameworks, robust prompting strategies, and interpretable optimization methods are essential for research advances to be successfully translated into reliable and effective systems for different stakeholders.

**Disclosure statement**

No potential conflict of interest was reported by the authors(s)

**Funding**

The author(s) reported that there is no funding associated with the work featured in the article.

**Author contributions**

All authors contributed equally to the conception, drafting, writing and editing of this manuscript. Each author has reviewed and approved the final text and has agreed to be responsible for the content and correctness of the work.

**Acknowledgement**

I would like to give credit to SciSpace for helping me to find related literatures and gap identification, and Grammarly for language improvement.